\documentstyle[a4]{article}

\begin{document}

\def\D{\Delta}
\def\R{I\!\!R}
\def\l{\lambda}
\def\D{\Delta}
\newcommand{\bib}{\bibitem}
\newcommand{\er}{\end{eqnarray}}
\newcommand{\br}{\begin{eqnarray}}
\newcommand{\be}{\begin{equation}}
\newcommand{\ee}{\end{equation}}
\newcommand{\epe}{\end{equation}}
\newcommand{\bea}{\begin{eqnarray}}
\newcommand{\bq}{\begin{eqnarray}}
\newcommand{\eea}{\end{eqnarray}}
\newcommand{\eq}{\end{eqnarray}}
\newcommand{\ba}{\begin{eqnarray}}
\newcommand{\ea}{\end{eqnarray}}
\newcommand{\epa}{\end{eqnarray}}
\newcommand{\ar}{\rightarrow}
\newcommand{\dslash}{\partial\!\!\!/}
\newcommand{\aslash}{a\!\!\!/}
\newcommand{\pslash}{p\!\!\!/}
\newcommand{\bslash}{b\!\!\!/}
\newcommand{\kslash}{k\!\!\!/}
\newcommand{\rslash}{r\!\!\!/}
\newcommand{\cslash}{c\!\!\!/}
\newcommand{\fslash}{f\!\!\!/}
\newcommand{\Dslash}{D\!\!\!\!/}
\newcommand{\Aslash}{{\cal A}\!\!\!\!/}
\def\r{\rho}
\def\d{\delta}
\def\T{\tilde{T}}
\def\k{\kappa}
\def\t{\tau}
\def\f{\phi}
\def\p{\psi}
\def\z{\zeta}
\def\ep{\epsilon}
\def\hx{\widehat{\xi}}
\def\na{\nabla}

\title{{\bf Dual embedding of the Lorentz-violating
electrodinamics and Batalin-Vilkovisky quantization.}}
\author{{\bf M. Botta Cantcheff$^{a}$},{\bf C.F.L. Godinho$^{a}$}, {\bf A.P. Ba\^{e}ta Scarpelli$^{b}$} \\
{\bf \thinspace and J.A. Helay\"el-Neto$^{a,b}\,$}
\thanks{{\tt E-mails:botta@cbpf.br, godinho@cbpf.br, scarp@fisica.ufmg.br,
helayel@cbpf.br.}}\vspace{2mm} \\
{\small {\bf $^{a}$}CBPF, Centro Brasileiro de Pesquisas F\'{\i }sicas }\\
{\small Rua Xavier Sigaud 150, 22290-180 Urca} \\
{\small Rio de Janeiro, RJ, Brazil.}\vspace{2mm}\\
{\small {\bf $^{b}$}Grupo de F\'{\i}sica Te\'{o}rica  Jos\'{e} Leite Lopes}\\
{\small Petr\'{o}polis, RJ, Brazil.}\vspace{2mm}.}
\maketitle

\begin{abstract}
\noindent
Modifications of the electromagnetic Maxwell Lagrangian in four
dimensions have been considered by some authors \cite{jac}. One
may include an explicit massive term (Proca) and a topological but
not Lorentz-invariant term within certain observational limits.
  We find the dual-corresponding gauge invariant version of this theory
by using the recently suggested gauge embedding method. We enforce
this dualisation procedure by showing that, in many cases, this is
actually a constructive method to find a sort of parent action,
which manifestly establishes duality. We also use the gauge
invariant version of this theory to formulate a Batalin-Vilkovisky
quantization and present a detailed discussion on the excitation
spectrum.

\end{abstract}

\section{Introduction}

It has been considered by  Carroll, Field and Jackiw, the
possibility of modifying the electromagnetic Maxwell Lagrangian to
include, for instance, explicit mass term and a topological term
\cite{jac},
\be
\label{e}
L_{\mu}[A_{\mu}]\,=-{\beta \over 4} F_{\mu \nu}F^{\mu
\nu}+\,\frac{m^2}{2}A_{\mu}A^{\mu}\,- \,L_{CS}[A_{\mu}]
\ee
where the $L_{CS}$ is a 3+1 version of the CS action, which
couples the dual electromagnetic tensor to an external four vector
$p$:
\be L_{CS}[A_{\mu}]=
-\,\frac{1}{4}p_{\alpha}A_{\beta}\ep^{\alpha\beta \mu\nu}F_{\mu
\nu} .
\ee

As it has been analysed, if this vector is fixed
to be (say) covariantly constant, $L_{CS}$ is gauge invariant but
not Lorentz invariant.  The experimental limits to the variations
of Maxwell model are also in ref. \cite{jac}.

This paper has a two-fold purpose: to construct a gauge invariant
version of the Maxwell modified theory via the gauging iterative
Noether method and to perform Batalin-Vilkovisky quantization; and
furthermore, to discuss on some properties of this procedure of
dualization; for instance, its relation to the parent action
approach.

Recently the so-called gauging iterative Noether Dualization
Method (NDM) \cite{w0} has been shown to be effective in
establishing some dualities between models \cite{w}. This method
is based on the traditional idea of a local lifting of a global
symmetry and may be realized by an iterative embedding Noether
counter terms. However, this method provides  a strong
suggestion of duality since it has been shown to give the expected
result in the paradigmatic duality between the so-called Self-Dual
model and Maxwell-Chern-Simons in three dimensions (SD-MCS). This
well-known correspondence was first established in detail by Deser
and Jackiw \cite{DJ}, and may be shown using a parent action approach \cite
{suecos}.

In this work we argue in favor of this
this method (NDM), since we show that, in certain cases, this may
actually be seen as a constructive
procedure to find a parent action; which would constitute a manifest
proof for the duality.

\section{The construction of the gauge invariant Lagrangian.}

Let us consider the massive Maxwell-Chern-Simons Lagrangian in
four dimensions \be \label{e1} L^{(0)}={- \beta \over 4} F_{\mu
\nu}F^{\mu \nu}+
\,\frac{m^2}{2}A_{\mu}A^{\mu}\,-\,\frac{1}{4}p_{\alpha}A_{\beta}\ep^{\alpha
\beta \mu\nu}F_{\mu \nu}. \ee Next, we are going to localize the
gauge symmetry, \be A_{\alpha}\to
A_{\alpha}\,+\,\partial_{\alpha}\eta , ~~~~~~~~~\delta
A_{\alpha}=\partial_{\alpha}\eta, \ee and calculate the first
variation of this lagrangian, \bea \delta L^{(0)}[A_{\mu}]\,=
\,\left( m^2 A_{\mu}+ \beta (\partial ^\nu F_{\nu \mu})-
\ep_{\alpha \beta \nu \mu}p^\alpha (\partial^\beta
A^\nu)\right)\partial^\mu \eta. \eea We may recognize the Noether
current as \be \label{e2} J_{\mu}\,=\,m^2 A_{\mu}+ \beta (\partial
^\nu F_{\nu \mu})- \ep_{\alpha \beta \nu \mu}p^\alpha
(\partial^\beta A^\nu) \ee so that we construct the first iterated
lagrangian by introducing an ancillary field $B$,
$L^{(1)}\,=\,L^{(0)}\,-\,JB$. If we admit now that $B$ transforms
according to \be \delta \, B_{\mu}\,=\,\delta\,A_{\mu} =
\partial_{\mu} \eta,
\ee then \be \label{e3}
\delta\,L^{(1)}\,=\,-(\delta\,J_{\mu})\,B^{\mu}. \ee In the other
hand we have \be \label{e4}
\delta\,J_{\mu}\,=\,m^2\delta\,A_{\mu}\,=\,m^2(\partial_{\mu}
\eta). \ee Note that by defining the second iterated lagrangian by
\be
 L^{(2)}\,=\,L^{(1)}\,+\,\frac{m^2}{2}\,B_{\mu}B^\mu,
\ee
and using (\ref{e3}) and (\ref{e4}), we get that the total
variation vanishes, $\delta L^{(2)}\,=\,0$. Let us write down the
explicit form of this action:
\bea
\label{e5}
L^{(2)}\,&=&{- \beta \over 4} F_{\mu \nu}F^{\mu
\nu}+\,\frac{m^2}{2}A_{\mu}A^{\mu}\, -
\,\frac{1}{4}p_{\alpha}A_{\beta}\ep^{\alpha\beta
\mu\nu}F_{\mu
\nu}  \nonumber  \\
&-&\left( m^2 A_{\mu}+ \beta (\partial ^\nu F_{\nu \mu})-
\ep_{\alpha \beta \nu \mu}p^\alpha (\partial^\beta A^\nu)
 \right)\,B^{\mu}\,+\,\frac{m^2}{2}\,B_\mu B^\mu.
\eea
Solving for $B$ we get the equation of motion
\be
-J + m^2 B=0
\ee
Plugging this back into (\ref{e5}), we obtain the remarkable gauge
invariant theory :
\bea
\label{e6}
L&=& {\beta \over 4 } F_{\mu \nu}F^{\mu \nu} +\frac 12 \ep_{\alpha \beta \nu \mu}p^\alpha (\partial^\beta A^\nu)A^ \mu -\frac {1}{2m^2} \left ( \ep_{\alpha \beta \nu \mu}p^\alpha (\partial^\beta A^\nu)\right)^2 \nonumber \\
&-& \frac {\beta ^2}{2m^2}(\partial _\mu F^{\mu \nu})^2 +\frac{\beta}{m^2}
\ep_{\alpha \beta \nu \mu}p^\alpha (\partial^\beta A^\nu)(\partial_ \rho F^{\rho \mu}).
\eea
We should observe also that the
massive exitation has been transformed in topological mass by this
procedure.

\subsection{On the Noether Dualization Method.}

Note the Maxwell term remain unafected under dualization and this
could be not considered ($\beta=0$).
This seems to be a characteristic of gauge invariant terms in the
initial action.
Notice further that
in the particular limit $\beta=0$, the action \ref{e5} is, in certain
sense,
a parent action for this duality, this is a frecuent
behavior of this dualization method.
To prove this, we
only must note that the last iterated lagrangian (\ref{e5}) is
precisely the so-called {\it parent action}. In other words, we
must show that by varying this with respect, first to $A$ and
second to $B$, the actions (\ref{e6}) and (\ref{e1}) are recovered.

Varying $L^2$ with respect to $B$, by construction \footnote{In
fact, this constitutes the final step of the Gauging Noether
method.}, equation \ref{e6} is obtained. Then if we variate this
action with respect to $A$, we obtain the equation of motion
\be
\label{e7} J(A-B)=0.
\ee
Next, let us define the new variable ${\tilde A} \equiv A-B$;
equation (\ref{e7}) corresponds to an action $L^{(0)}[{\tilde
A}]$. This completes our argument.

The similarity of this duality with the one in three dimensions between
the Self-Dual
model (SD) and Maxwell-CS (MCS) must to be remmarked. This is not
surprissing since, if we choice
the external vector $p_\mu$ to coincide with an (espacial) element of
cartesian basis
of the space-time, and writing the fields in components, one may
directly verify that
action (with $k=0$) reduces to SD in {\it three} dimensions and the
action (\ref{e6})
coincides with MCS. Thus, of course, the duality is preserved * .

Notice that this argument can be repeated in each case studied by
this Noether method, except when the non-linearity is marked, such
as in the case of the duality between Non-Abelian Self-Dual and
Yang-Mills
Chern Simons theories (in three dimensions). In these cases
another proof must be given, a suggestion to fill this gap was
given precisely for this problem in ref. \cite{par3d}

\section{Batalin-Vilkovisky Quantization.}

After the model has been analysed in terms of its particle
excitations and its consistency has been ascertained, our goal is
to implement a procedure to obtain a gauge fixed Generating
Functional of the new gauge invariant Lagrangian. For this reason
we will work with the field antifield formalism.
\newline
The standard form of field antifield quantization or BV scheme
\cite{BV} is based on imposing explicit BRST invariance of the
vacuum functional, by considering a closed equation that generates
the gauge algebra and uses it afterwards to construct the
functional generator of theory.  An important feature of the
standard BV is the anticanonical form of its phase space, where
the coordinates has a relationship conjugated with respect to the
antibracket operation.  Another important aspect of the field
antifield formalism is that canonical transformations
\cite{BV,ABG} do not change the form of the antibracket. At
classical level the action is constructed with the same Lagrangian
that one considered on the equation (\ref{e6}): \bea
L&=& {\beta \over 4 } F_{\mu \nu}F^{\mu \nu} +\frac 12 \ep_{\alpha \beta \nu \mu}p^\alpha (\partial^\beta A^\nu)A^ \mu -\frac {1}{2m^2} \left ( \ep_{\alpha \beta \nu \mu}p^\alpha (\partial^\beta A^\nu)\right)^2 \nonumber \\
&-& \frac {\beta ^2}{2m^2}(\partial _\mu F^{\mu \nu})^2
+\frac{\beta}{m^2} \ep_{\alpha \beta \nu \mu}p^\alpha
(\partial^\beta A^\nu)(\partial_ \rho F^{\rho \mu}), \nonumber
\eea where the gauge symmetry is \be \delta A_\mu\,=\,\partial_\mu
\eta \ee and the corresponding BRST symmetry can be read by \bea
sA_\mu&=&\partial_\mu C \\
sC&=&b. \eea Then we can deal now with the non  minimal  solution
of Master equation, by including the auxiliary fields
($b\,\,,\bar{C}$) \be S\,=\,\int d^4x \left( L\,+\,A_\mu^*
\partial^{\mu} C\,+\,\bar{C}^*b \right)
\ee

Notice that formally, the terms that breaks the Lorentz
invariance, more precisely the external vector $p$, can be treated
as an external source in order to perform a BV quantization of the
theory. However, since this is not a field of the non minimal
sector of theory its nature is BRST invariant. Therefore the path
integral is independent of the gauge choice, and its introduction
does not introduce subsequent modifications in the BV scheme to
the quantization procedure.   We are ready now to gauge fix the
model.  With this aim we will consider the functional fermion, \be
\Psi\,=\,\int d^4x \bar{C} \left(
\partial^\mu A_\mu\,-\,\frac{1}{2}b \right),
\ee the BRST variation of it is written by \be s\Psi\,=\,\int d^4x
\,\,\left ) b\partial^\mu
A_\mu\,-\,\bar{C}\partial^\mu\partial_\mu C\,-\,\frac{1}{2}b^2
\right ). \ee The generator functional  can be read with the gauge
fixed action by \be Z[\, p , \, J]\,=\,\int
DA_\mu\,DC\,D\bar{C}\,Db\,DA_\mu^*\,D\bar{C^*}\,\,
e^{-i\left(S\,+\,s\Psi + \, J_\mu A^\mu \right)}\, \ee After
functional integration over $A_\mu^*$, $\bar{C^*}$ and $b$ we are
able to write the final form for the generating functional \be
Z[\, p , \, J]\,=\, \int DA_\mu\,DC\, e^{\,-i\,\int d^4x
\left(L\,+\,\partial_\mu b
A^\mu\,+\,\partial_\mu\bar{C}\partial^\mu C + \, J_\mu A^\mu
\right)}, \ee where $L$ is the same written on (\ref{e6}).

\section{The Spectrum of the Theory and Unitarity Requirements.}

We now intend to investigate the pole structure of the gauge propagator
stemming from our Lagrangian
\br
L&=& {\beta \over 4 } F_{\mu \nu}F^{\mu \nu} +\frac 12 \ep_{\alpha \beta \nu \mu}p^\alpha (\partial^\beta A^\nu)A^ \mu -\frac {1}{2m^2} \left ( \ep_{\alpha \beta \nu \mu}p^\alpha (\partial^\beta A^\nu)\right)^2 \nonumber \\
&-& \frac {\beta ^2}{2m^2}(\partial _\mu F^{\mu \nu})^2 +\frac{\beta}{m^2}
\ep_{\alpha \beta \nu \mu}p^\alpha (\partial^\beta A^\nu)(\partial_ \rho F^{\rho \mu})
-\frac{1}{2\alpha}(\partial _\mu A^\mu),
\er
in which we have added a gauge-fixing term. By means of partial integrations in the action, we
can rearrange the Lagrangean in terms of spin operators
\br
L&=&\frac 12 A^{\mu}\left \{\left[-\beta \Box -\frac {1}{m^2}(p^2 \Box-
\lambda ^2)-\frac{\beta ^2 \Box ^2}{m^2}\right]\theta_{\mu \nu}+
\left(\frac{\Box}{\alpha}+\frac{\lambda ^2}
{m^2}\right)\omega _{\mu \nu}+\left (-1-\frac {2 \beta \Box}{m^2}\right)
 S_{\mu \nu} \right. \nonumber \\
&+&\left. \frac {\Box}{m^2}\Lambda_{\mu \nu}-\frac{\lambda}{m^2}
(\Sigma _{\mu \nu}+\Sigma_{\nu \mu})\right\}=\frac 12 A^{\mu}{\cal O}_{\mu \nu}A^{\nu},
\er
with $\theta_{\mu \nu}=g_{\mu \nu}-\frac {\partial _\mu \partial _\nu}{\Box}$
and $\omega_{\mu \nu}=\frac {\partial _\mu \partial _\nu}{\Box}$ being the transversal
and the longitudinal operators, respectively, and
\br
S_{\mu \nu}&=&\epsilon _{\mu \nu \alpha \beta}p^\alpha \partial ^\beta,   \\
\Lambda _{\mu \nu}&=& p_\mu p_\nu \,\,\,\,\,\,\mbox{and}  \\
\Sigma _{\mu \nu}&=& p_\mu \partial _\nu,
\er
generated by the inclusion of the external vector $p^\mu$. The $\lambda$ is just
$\Sigma _\mu \,\,^\mu=p_\mu \partial ^\mu$. The algebra of these operators
is shown in Table 1:
\begin{center}
\begin{tabular}{|c|c|c|c|c|c|c|}
\hline
& $\theta _{\,\,\,\,\,\nu }^{\alpha }$ & $\omega _{\,\,\,\,\,\nu }^{\alpha }$
& $S_{\,\,\,\,\,\nu }^{\alpha }$ & $\Lambda _{\,\,\,\,\,\nu }^{\alpha }$ & $%
\Sigma _{\,\,\,\,\,\nu }^{\alpha }$ & $\Sigma_\nu ^{\,\,\,\,\,\alpha}$ \\
\hline
$\theta _{\mu \alpha }$ & $\theta _{\mu \nu }$ & $0$ & $S_{\mu \nu }$ & $%
\Lambda _{\mu \nu }-\frac{\lambda }{\Box }\Sigma _{\nu \mu }$ & $\Sigma
_{\mu \nu }-\lambda \omega _{\mu \nu }$ & $0$ \\ \hline
$\omega _{\mu \alpha }$ & $0$ & $\omega _{\mu \nu }$ & $0$ & $\frac{\lambda
}{\Box }\Sigma _{\nu \mu }$ & $\lambda \omega _{\mu \nu }$ & $\Box_{\nu
\mu}$ \\ \hline
$S_{\mu \alpha }$ & $S_{\mu \nu }$ & $0$ & $f_{\mu \nu }$ & $0$ & $0$ & $0$
\\ \hline
$\Lambda _{\mu \alpha }$ & $\Lambda _{\mu \nu }-\frac{\lambda }{\Box }%
\Sigma_{\mu \nu }$ & $\frac{\lambda }{\Box }\Sigma _{\mu \nu }$ & $0$ & $%
v^{2}\Lambda _{\mu \nu }$ & $v^{2}\Sigma _{\mu \nu }$ & $\lambda
\Lambda_{\mu \nu}$ \\ \hline
$\Sigma _{\mu \alpha }$ & $0$ & $\Sigma _{\mu \nu }$ & $0$ & $\lambda
\Lambda _{\mu \nu }$ & $\lambda \Sigma _{\mu \nu }$ & $\Lambda_{\mu \nu}
\Box$ \\ \hline
$\Sigma_{\alpha \mu}$ & $\Sigma_{\nu \mu} -\lambda \omega_{\mu \nu}$ & $%
\lambda \omega_{\mu \nu}$ & $0$ & $v^2 \Sigma_{\nu \mu}$ & $v^2 \Box
\omega_{\mu \nu}$ & $\lambda \Sigma_{\nu \mu}$ \\ \hline
\end{tabular}
\vspace{2mm}

Table 1: Multiplicative Table fulfilled by $\theta ,\omega ,\,S,\,\Lambda $
and $\Sigma$. The products are supposed to obey the order ''row times
column''.
\end{center}
Making use of Table 1, we can invert ${\cal O}_{\mu \nu}$, which yelds  the following
vector propagator in momenta space:
\br
\langle A_\mu A_\nu \rangle &=&  \frac {i}{D}\left \{
\left[-m^2\left(k^2(k^2-m^2)-p^2k^2+(p \cdot k)^2\right) \right]\theta_{\mu \nu}\right. \nonumber \\
&+& \left. \left[-\frac{\alpha D}{k^2}-\frac {m^2(p \cdot k)^2 F}{k^2(k^2-m^2)}
\right]\omega _{\mu \nu}-im^2(2k^2-m^2)S_{\mu \nu} \right. \nonumber \\
&-& \left. \frac {m^2 F}{(k^2-m^2)}
\Lambda _{\mu \nu}+\frac {m^2(p \cdot k)F}{k^2(k^2-m^2)}(\Sigma _{\mu \nu}+ \Sigma _{\nu \mu})\right\}.
\er
The factor $F$ and the denominator $D$ are given by
\be
F=3k^2(k^2-m^2)+m^4+p^2k^2-(p \cdot k)^2
\ee
and
\be
D=\left[ k^4+p^2 k^2-(p \cdot k)^2 \right]\left[(k^2-m^2)^2+p^2 k^2-(p \cdot k)^2 \right].
\ee
In the above expressions, we have made $\beta=1$, in order to compare our spectrum with the obtained in the work \cite{AHBH}, where the original theory (the Lagrangian given by eq. (1) was analised. It is interesting to note that our denominator is simply the
product of two denominators: the one that comes from the original massive theory (eq.(1)) and
the one that comes from the non massive theory (eq.(1) without the Proca term). We have now  doubled the number of degrees of freedom, since the new Lagrangean has derivatives of higher order. This will be analised further.

Here we will analise situations in which the external vector $p^\mu$ is purely space-like. The
time- and light-like terms allows for non physical poles in the theory of eq(1) as it is shown in ref. \cite{AHBH}. Then we take $p^\mu=(0,0,0,\mu)$ and $k^\mu=(k^0,0,0,k^3)$. The denominator, then, reads
\be
D=[k^2-m_1'^2][k^2-m_2'^2][k^2-m_1^2][k^2-m_2^2],
\ee
with the poles
\bq
k_0^2&=&m_1^2=\frac 12 \left [2(m^2+k_3^2)+\mu^2+\mu \sqrt{\mu^2+4(m^2+k_3^2)}\right], \\
k_0^2&=&m_2^2=\frac 12 \left [2(m^2+k_3^2)+\mu^2-\mu \sqrt{\mu^2+4(m^2+k_3^2)}\right], \\
k_0^2&=&m_1'^2=\frac 12 \left [2k_3^2+\mu^2+\mu \sqrt{\mu^2+4k_3^2}\right], \\
k_0^2&=&m_2'^2=\frac 12 \left [2k_3^2+\mu^2-\mu \sqrt{\mu^2+4k_3^2}\right].
\eq
Besides these poles, we still have the poles $k_0^2=k_3^2$ ($k^2=0$) and $k_0^2=m^2+k_3^2$
($k^2=m^2$). The poles $m_1^2$, $m_2^2$ and $m^2+k_3^2$ come from the massive sector, while the other three arise from the massless sector.

We start off with the calculus of the residue matrix for the pole $k_0^2=m_1^1$, for which we
find
\be
R_{1}=\frac{1}{\sqrt{\mu^2+4(m^2+k_3^2)}}\left(
\begin{array}{cccc}
0& 0 & 0 & 0 \\
0 & m_1 & i m_1 & 0 \\
0 & - i m_1  & m_1  & 0 \\
0 & 0 & 0 & 0
\end{array}
\right),
\ee
with the eigenvalues
\br
\lambda _1 &=& 0  \\
\lambda _2 &=& 0  \\
\lambda _3 &=& \frac{2|m_1|}{\sqrt{\mu^2+4(m^2+k_3^2)}}>0          \,\,\,\,\,\, \mbox{and} \\
\lambda _4 &=& 0.
\er
As it can be seen, this pole is to be associated with only one physical degree of freedom, since we
have only one non-null eigenvalue. The same hapens with the pole $k^2=m_2^2$, for which the
results are completely similar with the exchange of $m_1$ for $m_2$.

We now study the pole $k_0^2=m^2+k_3^2$. The associated  residue matrix reads as below
\begin{equation}
R_{m}=\left(
\begin{array}{cccc}
\frac{k_{3}^{2}}{m^2}  & 0 & 0 &
\frac{\left| k_{3}\right| \left( m^{2}+k_{3}^{2}\right) ^{1/2}} {m^2}\\
0 & 0 & 0 & 0 \\
0 & 0 & 0 & 0 \\
\frac{\left| k_{3}\right| \left( m^{2}+k_{3}^{2}\right) ^{1/2}}{m^2} & 0 & 0 & \frac{\left( m^{2}+k_{3}^{2}\right)}{m^2},
\end{array}\right)
\end{equation}
with a unique non-vanishing eigenvalue:
\be
\lambda=\frac{m^2+2k_3^2}{m^2}>0.
\ee
These results obtained for the three initial poles, coming from the original theory, that includes a Proca term, are exactly the same obtained in \cite{AHBH}, where the Lagrangean
eq(1) was studied. They express three degrees of freedon, which one coming from one pole.

The study of the three remaining poles brings about an interesting result: the residues found for the poles coming from the massless factor of $D$ are simply the residues found for the massless theory up to  a minus sign. This suggest that maybe the total propagator can be rearranged so as to be expressed as the difference between the
propagators of the massive and the non-massive theories. Indeed, we have:
\be
\langle A_\mu A_\nu \rangle=\langle A_\mu A_\nu \rangle _1-\langle A_\mu A_\nu \rangle _2
-\frac 1{m^2} \omega_{\mu \nu},
\ee
with
\begin{eqnarray}
\langle A_{\mu }A_{\nu }\rangle _1&=&\frac{i}{D_1}\left\{ -(k^{2}-m^{2})\theta
_{\mu \nu }-\frac{(p\cdot k)^{2}}{%
(k^{2}-m^{2})} \omega _{\mu \nu }\right.  \nonumber \\
&&\left. -i S_{\mu \nu }-\frac{k^{2}}{(k^{2}-m^{2})}\Lambda _{\mu
\nu }+\frac{(p\cdot k)}{(k^{2}-m^{2})}\left( \Sigma _{\mu \nu
}+\Sigma _{\nu \mu }\right) \right\}   \label{13}
\end{eqnarray}
and
\begin{eqnarray}
\langle A_{\mu }A_{\nu }\rangle _2 &=&\frac{i}{D_2}\left\{ -k^{2}\theta
_{\mu \nu }-\frac{(p\cdot k)^{2}}{
k^{2}} \omega _{\mu \nu }\right.  \nonumber \\
&&\left. -i S_{\mu \nu }-\Lambda _{\mu
\nu }+\frac{(p\cdot k)}{k^{2}}\left( \Sigma _{\mu \nu
}+\Sigma _{\nu \mu }\right) \right\} ,  \label{13}
\end{eqnarray}
where we have made the gauge choice $\alpha=0$. This is a remarkable result. The minus sign in the second propagator guarantees us that the physical particles predicted by the original theory
are the same in its dual theory, since the negative eigenvalus obtained in the second part are compatible with negative norm states. This apparently surprising result may actually be traced
back to the case of the usual Proca theory. If we proceed to the dualisation of the action for
the Proca field, as we have done in Section 1 (without the Lorentz-breaking term), we can show
that the $A_\mu$-propagator of the dual theory also turns out to have the same relation with the massive and the massless $A_\mu$-propagators.

\section{A Final Comment.}

The gauge invariant Maxwell Theory was modified in such a form
that the gauge invariance is broken within certain observational
limits. We showed here that this invariance can be restored and
this modified version of the electromagnetism results to be
equivalent to a novel gauge invariant theory (equation
(\ref{e6})). However the Lorentz invariance remains broken.

{\bf Acknowledgements}:  The authors express their gratitude to
CNPq, for the invaluable financial help. M.B.C was supported by
CLAF.


\begin{thebibliography}{99}

\bibitem{jac} S. M. Carrol, George B. Field and Roman Jackiw,
Phys. Rev. D 41 (1990) 1231

\bibitem{w0} M. A. Anacleto, A. Ilha, J. R. S. Nascimento, R. F.
 Ribeiro and C. Wotzasek, Phys. Lett. B504 (2001) 268.

\bibitem {w} A. Ilha and C. Wotzasek, Nucl.Phys. B604 (2001) 426.

\bibitem{DJ} S. Deser and R. Jackiw, Phys. Lett. B 139
(1984) 2366.

\bibitem{suecos} For a recent review in the use of the master action
 in proving duality in diverse areas see: S. E. Hjelmeland, U.
Lindstr\"om, UIO-PHYS-97-03,
 May 1997. e-Print Archive: hep-th/9705122

\bibitem{par3d} M. Botta Cantcheff, Phys.Lett. B528 (2002) 283-287.

\bibitem{AHBH} A. P. Ba\^eta Scarpelli, H. Belich, J. L. Boldo and J. A. Helayel-Neto,
{\it Aspects of causality and unitarity and comments on vortex-like configurations
in an abelian model with a Lorentz-breaking term}, hep-th/0204232 (to appear in Phys.
Rev. D)

\bibitem{BV} I. A. Batalin and G. A. Vilkovisky Phys. Lett. B 102
(1981) 27

\bibitem{ABG} E. M. C. Abreu, N. R. F. Braga and C. F. L. Godinho,
Nuc. Phys (1987)



\bibitem{TPvN} P. K. Townsend, K. Pilch and P. van Nieuwenhuizen, Phys.
Lett. B 136 (1984) 38.

\bibitem{chap1} E. Fradkin and F. A. Schaposnik, Phys. Lett.
B338 (1994) 253; E. Fradkin, F. A. Schaposnik,  Phys. Lett.B358
(1994) 253 .



\end{thebibliography}
\end{document}